# Compensating effect of solenoids with quadrupole lenses

Vladimir N Litvinenko[1] and Alexander A. Zholents[2]

[1] *Department of Physics and Astronomy, Stony Brook University, Stony Brook, NY*

[2] *Argonne National Laboratory, Lemont, IL*

*Abstract.* In this paper we present methods for compensating influence of solenoids, including the de-coupling of transverse motion, using quadrupole lenses. This technique could be used for variety of schemes in storage rings and transport channels where solenoids are required – for example, in storage rings operating with longitudinally polarized beams.

We developed a number of methods when betatron motion is coupled only locally (e.g. coupling exists only in the insert) while the rest of the storage ring optics remains undisturbed. Such inserts would allow engendering longitudinal polarization of colliding beams in existing colliders.

We present a set of selected simulations for inserts removing the coupling introduced by solenoids while maintaining the matching with the optics of the rest if the storage ring.

*Foreword:* This paper was originally written in 1981 and published in Russian as a preprint at Novosibirsk Institute of Nuclear Physics (INP) [1], where authors were working at the time. During this period scientist at INP were actively discussing possibility of longitudinal polarization in VEPP-4 electron-position collider. It was recognized that introducing solenoid flipping vertical spin by 180 degrees would result in fully coupled and unstable transverse motion. Our paper was addressing this problem and finding a range of possible solutions. In 1984 the preprint was partially translated in English at DESY by I. Schulz-Dahlen and published as DESY Internal Report [2][1]. Meanwhile, our compensation scheme proven to be rather popular and has been employed in number of accelerator designs. Some implementations of our proposed schemes can be found in [3-24]. Ideas originated from our proposed method found applications not only in beam polarization schemes but also in beam manipulations for magnetized electron cooling and FELs. As can be seen from these references, this 36-year old scheme remains popular. This is the main reason why we decided, finally, to publish it as regular journal article. We redraw the figures, retyped formulae to fit modern standards, re-fit the solutions while keeping the scientific content as well as original references [25-30] as in the original paper [1].

---

[1] *Authors learned about this unauthorized translation many years later. In addition one of our last names being misspelled, there were few additional typos, which most likely originated from misreading our handwritten equations.*

**Introduction**

A number of schemes proposed for generating longitudinal beam polarization in electron-positron based on use of strong solenoids rotating particle's spin for angles of $\pi/2$ or $\pi$ [25]. To rotate spin of electron (or positron) on angle $\varphi$ a solenoid with filed integral of

$$\frac{e}{pc}\int B_s ds = \varphi \qquad (1)$$

is required, where $e$ is the charge of electron (position), $p$ is the particle momentum, $c$ is the speed of the light, and $B_s$ is the solenoidal field[2]. In practical units it writes as:

$$\int B_s ds \approx 10[Tm] \cdot \frac{\varphi}{\pi} \cdot E[GeV] \qquad (2)$$

for ultra-relativistic particles with $E \cong pc$. Such solenoids would introduce a strong coupling between horizontal and vertical betatron oscillations and also could significantly affect optical functions of the collider. In spite of these challenges, the advantages of using solenoids for attaining longitudinal polarization in collider are very significant. For example, such schemes do not require modifications of the ring geometry can (in principle) operate at any energy and can be just inserted into a straight section of a collider. It makes such scheme very attractive for existing colliders.

Our schemes provide for localized compensation of the solenoid influence on beam dynamics without affecting beam polarization. Hence, longitudinal polarization in colliders can be attained relatively straightforward using solenoids.

A compensation scheme based on four SQ-quadrupole[3] lenses was proposed by A. M. Kondratenko for a specific case of solenoid rotating spin by $\varphi = \pi$. This scheme, described in detail in Ref. [26], cancels transverse coupling but does not address additional focusing introduced by the solenoid. Authors of [26] propose to use additional eight quadrupoles to address this problem – a rather cumbersome solution, which could complicate machine operation.

In this paper, we present a systematic approach of designing compensation schemes providing for localization of the effects introduced by a solenoid. We present a general case of local compensating influence on beam dynamics of solenoid rotating spin on an arbitrary angle $\varphi$. It also can be used during the ramping of the solenoid's field. Hence, our scheme could be used for providing longitudinal polarization of colliding beams in VEPP-4 or any other collider.

**I. Transport matrices of solenoid and a rotated quadrupole.**

It is well known (see for example [27]) that magnetic field of solenoid is fully and uniquely described by magnetic field on its axis:

---

[2] *Please note that we are using CGS units in this paper.*
[3] *E.g. a quadrupole turned by 45 degrees around its axis.*

$$B_s = B_s(s), \tag{3}$$

with *s* being the coordinate along the solenoid axis. In this case, the EM vector potential has only one component and can be explicitly written as function of $B_s$ and its derivatives:

$$A_\varphi = \sum_{n=0}^{\infty} \frac{(-1)^n}{n!(n+1)!} B_s^{(2n)} \cdot \left(\frac{r}{2}\right)^{2n+1}; \quad B_s^{(n)} \equiv \frac{\partial^n B_s(s)}{\partial s^n}, \tag{4}$$

where we are using cylindrical coordinates $(s,r,\varphi)$. Classical motion of a charge particle in the solenoid is fully described by its Lagrangian:

$$L = -mc^2\sqrt{1-\frac{\vec{v}^2}{c^2}} + \frac{e}{c}A_\varphi \cdot r \cdot \dot\varphi; \quad \vec{v} \equiv \frac{d\vec{r}}{dt}; \quad \dot\varphi = \frac{d\varphi}{dt}, \tag{5}$$

where *m* and *e* are mass and electric charge of the particle. It is well known that energy of a particle in the magnetic field is preserved

$$E = \gamma mc^2 = inv; \quad \gamma = \left(1-\frac{\vec{v}^2}{c^2}\right)^{-1/2} = inv, \tag{6}$$

and that axial (rotation) symmetry of the Lagrangian indicates existence of the integral of motion:

$$\frac{\partial L}{\partial \varphi} = 0 \Rightarrow P_\varphi = \frac{\partial L}{\partial \dot\varphi} = \gamma mr^2\dot\varphi + \frac{e}{c}A_\varphi \cdot r = inv = \gamma mr_o^2\dot\varphi_o; \tag{7}$$

where $r_o$ and $\dot\varphi_o$ are values that the solenoid entrance where $A_\varphi$=0. Thus, the solenoidal field causes additional rotation of particle's position

$$\dot\varphi = \frac{r_o^2}{r^2}\dot\varphi_o - \frac{e}{\gamma mc}\frac{A_\varphi}{r}; \quad \Delta\dot\varphi = -\frac{e}{\gamma mc}\frac{A_\varphi}{r}; \tag{8}$$

or changing independent variable to *s*:

$$\Delta\varphi' = -\frac{e}{p_s c}\frac{A_\varphi}{r}; \quad f' \equiv \frac{df}{ds} = \frac{1}{v_s}\frac{df}{dt}; \frac{ds}{dt} = v_s; \quad p_s = \gamma mv_s; \tag{9}$$

Since solenoid aberrations are off interest in this paper, we can use paraxial approximation when *r*-motion is decoupled from $\varphi$ and (9) can be rewritten as

$$\Delta\varphi'(s) \cong -\frac{eB_s(s)}{2pc} \equiv \frac{B_s(s)}{2\overline{B\rho}}; \tag{10}$$

where $\overline{B\rho} = \frac{pc}{e}$ is the rigidity of the beam frequently used in accelerator physics.

We demonstrated that coupling between vertical and horizontal motion introduced by solenoidal field can be compensated using quadrupole lenses. Such compensation is required, for example, when solenoids are used for producing longitudinal polarization in storage rings. Indeed, in this case, the motion is completely couple and also unstable (e.g. trace of the 4x4 matrix is equal zero).

Traditionally transverse motion of particles is decoupled and it is a linear superposition of horizontal ($x$) and vertical ($y$) betatron oscillations. Since oscillation in a plane ($\vec{v}_\perp \| \vec{r}_\perp$, $[\vec{v}_\perp \times \vec{r}_\perp] = 0$) correspond to $\dot{\varphi}_o = 0$, and

$$\Delta\varphi = -\frac{e}{2pc}\int B_s(s)ds \equiv \frac{1}{2\overline{B\rho}}\int B_s(s)ds, \qquad (11)$$

is nothing else but rotation of the plane of oscillation. Hence, a solenoid would not couple $x$ and $y$ betatron oscillations when

$$\int B_s(s)ds = 2\pi n \cdot \overline{B\rho}, \ n = 1,2,3... \qquad (12)$$

e.g. when spin is rotated by $2\pi n$ in its original direction.

Equation of motions in solenoid are natural to be studied in rotating coordinate system. As shown in [28, 31-33][4], introducing torsion

$$\kappa(s) = -\frac{eB_s(s)}{2pc} = \frac{B_s(s)}{2\overline{B\rho}} \qquad (13)$$

decouples transverse motion from rotation and transforming equation of motion into a simple oscillator :

$$x'' + \kappa^2(s)x = 0; \ y'' + \kappa^2(s)y = 0; \qquad (14)$$

From here we conclude that 4x4 transport matrix of an arbitrary solenoid has following form:

$$\mathbf{M}_s = \mathbf{R}(\theta)\begin{pmatrix} \mathbf{F} & \mathbf{0} \\ \mathbf{0} & \mathbf{F} \end{pmatrix} \equiv \begin{pmatrix} \mathbf{F} & \mathbf{0} \\ \mathbf{0} & \mathbf{F} \end{pmatrix}\mathbf{R}(\theta); \ \mathbf{R}(\theta) = \begin{pmatrix} \mathbf{I}\cos\theta & \mathbf{I}\sin\theta \\ -\mathbf{I}\sin\theta & \mathbf{I}\cos\theta \end{pmatrix};$$

$$\mathbf{I} = \begin{pmatrix} 1 & 0 \\ 0 & 1 \end{pmatrix}; \ \mathbf{0} = \begin{pmatrix} 0 & 0 \\ 0 & 0 \end{pmatrix}; \ \theta = -\frac{e}{2pc}\int_0^{L_s} B_s(s)ds; \qquad (14)$$

where $\mathbf{F}$ is a 2x2 focusing matrix of solenoid satisfying

$$\frac{d\mathbf{F}(s)}{ds} = \begin{bmatrix} 0 & 1 \\ -\kappa^2(s) & 0 \end{bmatrix}\cdot \mathbf{F}(s); \ \mathbf{F}(0) = \mathbf{I}, \qquad (15)$$

$\mathbf{R}$ is a 4x4 rotation matrix, and $L_s$ is the length of solenoidal field. In a specific case of hard-edge solenoid ($B_s = B_o$, $0 \leq s \leq L_s$, and zero outside the range) (15) gives well known result [29]:

---

[4] In contrast with [28], Refs. [31-33] are in English and accessible from web

$$\mathbf{F} = \begin{bmatrix} \cos\theta & \frac{1}{\kappa}\sin\theta \\ -\kappa\sin\theta & \cos\theta \end{bmatrix}; \quad \kappa = \frac{eB_o}{2pc}. \quad (16)$$

In worth mentioning for the further use, that matrices of a solenoid and a drift commute with matrix of rotation, $\mathbf{R}^5$:

$$\mathbf{M}_S \mathbf{R} = \mathbf{R}\mathbf{M}_S; \quad \mathbf{M}_D \mathbf{R} = \mathbf{R}\mathbf{M}_D; \quad \mathbf{M}_D = \begin{pmatrix} \mathbf{M}_d & 0 \\ 0 & \mathbf{M}_d \end{pmatrix}; \quad \mathbf{M}_d = \begin{pmatrix} 1 & s_d \\ 0 & 1 \end{pmatrix}. \quad (17)$$

Transport matrix of a quadrupole rotated at angle α can be easily rotating the coordinate system by angle α, transporting particles through the normal quadrupole, and rotating coordinates back:

$$\mathbf{M}_{Q\alpha} = \mathbf{R}^{-1}(\alpha)\mathbf{M}_Q \mathbf{R}(\alpha); \quad \mathbf{M}_Q = \begin{pmatrix} \mathbf{M}_{Qx} & 0 \\ 0 & \mathbf{M}_{Qy} \end{pmatrix}; \quad \mathbf{R}^{-1}(\alpha) = \mathbf{R}(-\alpha) = \mathbf{R}^T(\alpha). \quad (18)$$

## II. Methods for designing solenoid compensation schemes.

Arbitrary transport 4x4 matrix is symplectic[6] and only 10 of its 16 elements are independent. Decoupling of transverse degrees of motion is sufficient to zero 4 out of these 10 parameters, e.g. to eliminate non-diagonal blocks. Thus, no more that four quadrupoles are required for the decoupling. Compensation for focusing introduced by solenoid one would need additional parameters, either additional quadrupoles or, possibly, change of their positions.

*Method 1.* It is convenient to start our description of compensation methods from scheme shown in Fig.1: two identical solenoids are separated by a straight section with six normal quadrupoles.

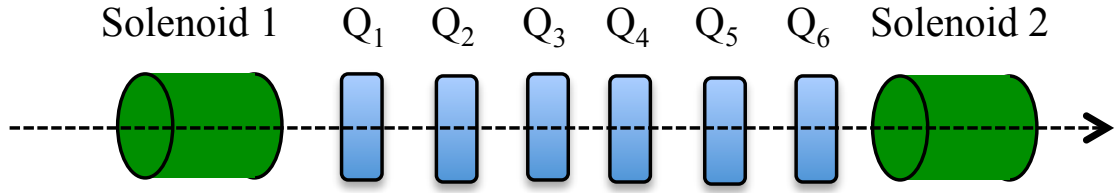

Fig. 1. One of possible solenoid compensation schemes with six quadrupoles.

---

[5] This is true for any block-diagonal matrix: $\mathbf{M}_d = \begin{pmatrix} \mathbf{A} & 0 \\ 0 & \mathbf{A} \end{pmatrix} \Rightarrow \mathbf{MR} = \mathbf{RM}$.

[6] We are using Hamiltonian description for the particles motion with canonical momenta differing from mechanical momenta: $\vec{P} = \vec{p} + \frac{e}{c}\vec{A}$

According to eqs. (14) and (17), the entire 4x4 transport matrix can be written as

$$\mathbf{M} = \mathbf{R} \begin{pmatrix} \mathbf{A} & 0 \\ 0 & \mathbf{B} \end{pmatrix} \mathbf{R}, \tag{19}$$

where **A** and **B** are some 2x2 symplectic matrices (see Appendix 1). It is easy to show (see Appendix 2) that for matrix **M** to be block diagonal it is necessary and sufficient that

$$\mathbf{B} = -\mathbf{A} \implies \mathbf{M} = \begin{pmatrix} \mathbf{A} & 0 \\ 0 & -\mathbf{A} \end{pmatrix}. \tag{20}$$

Since $\det \mathbf{A} = \det \mathbf{B} = 1$, condition (20) could be satisfied by 3 quadrupoles. Three remaining quadrupoles would be sufficient to bring **A** to any desirable symplectic value, for example to an unit matrix or matrix of a drift with the length equal to that of the full system, $l_{tot}$:

$$\mathbf{A}_u = \begin{pmatrix} 1 & 0 \\ 0 & 1 \end{pmatrix}; \quad \mathbf{A}_d = \begin{pmatrix} 1 & l_{tot} \\ 0 & 1 \end{pmatrix}. \tag{21}$$

The later will be a natural choice for an insertion replacing a drift section with length *l*. Since sign of the matrix does not affect $\beta$- and $\alpha$- functions, this would provide for optics match in both directions. Naturally, one of the tune would be changed by 1/2.

*Method 2.* Second, and significantly more general, compensation method is based on idea how to bring the transport matrix of an arbitrary combination of solenoids and quadrupole to the form shown in eq. (19). In this scheme, number of solenoids and quadrupoles can be arbitrary. The solution requires quadrupoles to be rotated at correct angle. As can be seen from eq. (14), transverse motion of in rotating coordinate system with $\kappa(s) = -eB_s(s)/2pc$ remains uncoupled. Hence, introducing quadrupole lens at any location *s* rotated on the angle

$$\theta(s) = \int_0^s \kappa(z) dz \tag{22}$$

would keep the motion uncoupled

$$x'' + (\kappa^2(s) + g(s))x = 0; \quad y'' + (\kappa^2(s) - g(s))y = 0; \tag{23}$$

where $g(s) = -\dfrac{eG(s)}{pc}$; $G = \dfrac{\partial B_y}{\partial x}$ is quadrupole focusing term.

It is obvious that in that rotated coordinate system the matrix of quadrupole is a block-diagonal. Thus, in the normal coordinate system the matrix would have following form:

$$\mathbf{M}^* = \mathbf{R}(\theta_o) \begin{pmatrix} \mathbf{A} & 0 \\ 0 & \mathbf{B} \end{pmatrix}; \quad \theta_o = \int_0^{l_{tot}} \kappa(z) dz,, \tag{24}$$

where $l_{tot}$ is the length of the system, and matrices **A** and **B** are simple ordered products of block-diagonal matrices $\mathbf{M}_{Qx}, \mathbf{M}_{Qy}, \mathbf{M}_d, \mathbf{F}$.

We can transfer matrix $\mathbf{M}^*$ to the form in eq. (19) by rotating all system by $-\theta_o/2$:

$$\mathbf{M} = \mathbf{R}\left(-\frac{\theta_o}{2}\right)\mathbf{M}^*\mathbf{R}\left(\frac{\theta_o}{2}\right) = \mathbf{R}\left(\frac{\theta_o}{2}\right)\begin{pmatrix} \mathbf{A} & 0 \\ 0 & \mathbf{B} \end{pmatrix}\mathbf{R}\left(\frac{\theta_o}{2}\right)., \quad (24)$$

Hence any system of quadrupoles and solenoids would have desirable form of eq. (19) if each quadrupole lens is rotated by angle

$$\theta_r(s) = \int_0^s \kappa(z)dz - \frac{\theta_o}{2} = \frac{1}{2}\left(\int_0^s \kappa(z)dz - \int_s^{l_{tot}} \kappa(z)dz\right) \quad (22)$$

where $s$ is the quadrupole's location. Naturally, all considerations about compensation strategy used for the first method are fully applicable to this case.

A simplest solenoid compensation scheme using second methods is shown in Fig.2.

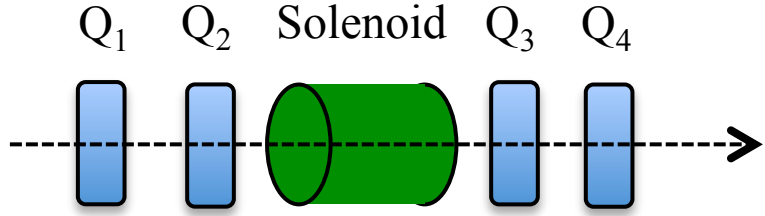

Fig. 2. A simplest decoupling scheme based on the Method 2, for a solenoid with spin rotation angle of $2\theta_o$. First quadrupole pair, $Q_1$ and $Q_2$, are rotated by angle of $-\theta_o/2$, while the second pair, $Q_3$ and $Q_4$, are rotated by angle $\theta_o/2$.

In this scheme solenoid (designed to rotate spin by angle $2\theta_o$) rotates planes of oscillations by $\theta_o$. Hence, two pairs of quadrupoles are rotated by angles $-\theta_o/2$ and $+\theta_o/2$ correspondingly. Such scheme allows decoupling transverse motion (e.g. satisfying condition (20)), but is insufficient to make a desirable values of matrix **A**. Authors of [26] described in detail specific case of such scheme for $\theta_o = \pi/2$.

More complicated schemes based on Method 2 are shown in Figs. 3 and 4.

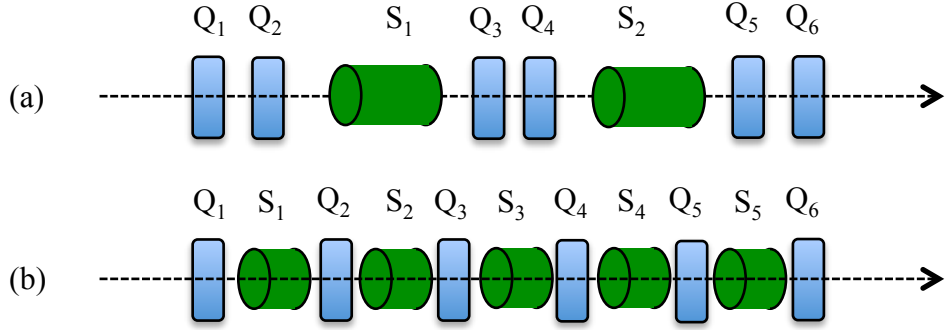

Fig. 3. Two possible compensation schemes based on Method 2 with six quadrupoles solenoid split into (a) two parts and (b) five parts. Six quadrupoles are typically sufficient for compensating effect of solenoids on both coupling and beam optics.

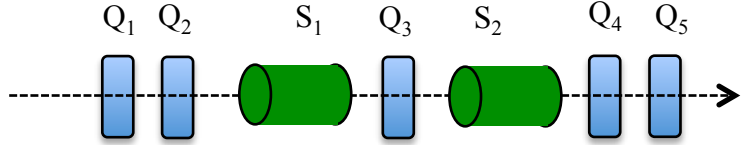

Fig. 4. A possible compensation scheme based on Method 2 with five quadrupoles and solenoid split into two parts and (b) five parts. Five quadrupoles are typically insufficient for arbitrary fitting, but work for bilaterally symmetric systems.

### III. Numerical examples.

An optimal choice of compensation scheme depends on specifics of the problem, accelerator geometry and hardware limitations. First, and foremost, one have to choose the most suitable from of matrix **A**. Here we assume that it is convenient to use either unit matrix $\begin{pmatrix} 1 & 0 \\ 0 & 1 \end{pmatrix}$ or that of the drift $\begin{pmatrix} 1 & l_{tot} \\ 0 & 1 \end{pmatrix}$ with the length of the insert. Naturally, some applications could require another type of matrix **A**.

As we mentioned before, the transport matrix of the insert in $y$ direction will be – **A**. Sign of the matrix does not affect fitting $\beta$- and $\alpha$- functions. But if the inset is installed into a straight section with non-zero horizontal dispersion, sign of the matrix would matter. In other words, horizontal directions should requires to be fit first and its fractional betatron tune would remain the same. Betatron tune in $y$ direction would change by a half[7].

We numerically simulated a number of compensation schemes to test soundness of our proposed methods. Optimization code OPTI [30] was used for this purpose.

---

[7] It is always possible to add additional three quadrupole lenses to match this system into an arbitrary lattice and to keep fractional betatron tunes unchanged. Here we only considering economic version of such compensation schemes.

We used the scheme shown in Fig. 1 to create a unit matrix in horizontal direction and minus unit in the vertical direction for an arbitrary rotation angle of solenoid. Unit transport matrices we utilized for straight section storage ring The parameters of the insert lattice are shown in Table 1, while Fig. 5 illustrates dependences of the quadrupole lenses gradients on the spin rotation angle of the insert [8].

Table 1. Lattice parameters of the solenoid insert shown in Fig.1.

| Parameter | Value | Comment |
|---|---|---|
| Beam energy, GeV | 3.0 | Electrons |
| Solenoid length, cm | 82 | Two units |
| Form of matrix A | Unit | |
| Method | 1 | |
| Quadrupole length, cm | 20 | Six quadrupoles |
| Drifts length between quadrupoles, cm | 180 | Five drift |
| Drifts between solenoid and quadrupole, cm | 8 | |
| Total length of insert, m | 12 | |

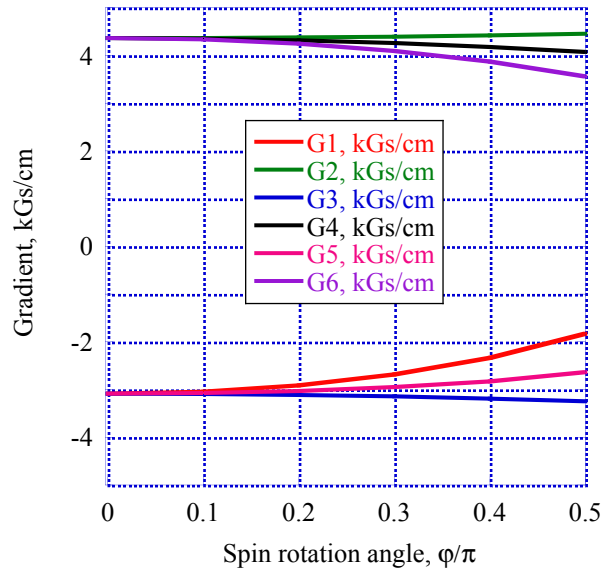

Fig.5. Dependence of the gradients in quadrupole lenses on the spin rotation angle. These gradients are required to maintain unit matrix in horizontal direction and minus unit in the vertical direction. A 1 kGs/cm field gradient of corresponds to 10 T/m in SI units.

[8] While the original optimization was done using code OPTI, to restore the exact numbers for fig. 5 we used a simple optimizer written in Mathematica [34].

Main advantage of this scheme that it allows changing strength of solenoids with and to adiabatically change spin direction in the presence of circulating beams in the collider. To avoid depolarizing resonance, it is advisable that such tuning would occur at half-integer spin tune $v_s = 1/2$.

Two other numerical results illustrate schemes using method 2. Table 2 lists lattice parameters and Table 3 quadrupole gradient for an scheme shown in Fig. 3 (b), with matrix $\mathbf{A} = \begin{pmatrix} 1 & l_{tot} \\ 0 & 1 \end{pmatrix}$ for the spin-flipping insert. Similarly, Tables 4 and 5 list parameters for the scheme shown in Fig. 4 and matrix of the drift with the length of full insert. It worth mentioning, that this scheme has only five quadrupoles and takes full advantage of bilateral symmetry and would require only 3 power supplies if Q1/Q5 and Q2/Q4 fed in series. It is also the shortest of three schemes we simulated with total length of 4.3 m.

Table 2. Lattice parameters of the solenoid insert shown in Fig.3(b).

| Parameter | Value | Comment |
|---|---|---|
| Beam energy, GeV | 2.0 | Electrons |
| Solenoid length, cm | 50 | Five solenoids |
| Solenoidal field, kGs | 83.78 | 8.378 T |
| Spin rotation angle | 180º | |
| Form of matrix A | Drift | The length of insert |
| Method | 2 | |
| Quadrupole length, cm | 30 | Six quadrupoles |
| Drifts between all elements, cm | 10 | |
| Total length of insert, m | 5.3 | |

Table 3. Values of quadrupole gradients and rotation angles for lattice in Table 2.

| Quadrupole | Q1 | Q2 | Q3 | Q4 | Q5 | Q6 |
|---|---|---|---|---|---|---|
| Gradient, kGs/cm | -0.331 | 0.089 | -0.349 | 3.826 | 1.532 | -1.103 |
| Rotation angle | -45º | -27º | -9º | 9º | 27º | 45º |

Table 4. Lattice parameters of the solenoid insert shown in Fig.4.

| Parameter | Value | Comment |
|---|---|---|
| Beam energy, GeV | 2.0 | Electrons |
| Solenoid length, cm | 125 | Two solenoids |
| Solenoidal field, kGs | 83.78 | 8.378 T |
| Spin rotation angle | 180º | |
| Form of matrix A | Drift | The length of insert |
| Method | 2 | |
| Number of quadrupoles | 5 | |
| Drifts between all elements, cm | 10 | |
| Total length of insert, m | 4.3 | |

Table 5. Parameter of quadrupoles and rotation angles for lattice in Table 4.

| Quadrupole | Q1 | Q2 | Q3 | Q4 | Q5 |
|---|---|---|---|---|---|
| Length, cm | 20 | 20 | 40 | 20 | 20 |
| Gradient, kGs/cm | -3.252 | 2.949 | -3.128 | 2.3949 | -3.252 |
| Rotation angle | -45° | -45° | 0° | 45° | 45° |

Overall, schemes using second method are effective by unitizing solenoids as effective drifts between quadrupoles, and therefore are generally more compact. At the same time, if spin rotation angle have to be adjustable, it would require not only changing quadrupoles strength abut also retaining quadrupoles (or at least their fields) according eq. (22).

## IV. Conclusions

In this paper we developed various scheme to compensate influence of solenoids proving longitudinal polarization electron-position colliders. Nevertheless, the proposed compensation schemes could be used for a variety of accelerator applications. For example, they can be used to compensate influence of detectors with strong solenoidal fields. This task usually solved by the use of compensating solenoids. Using quadrupole scheme shown in Fig. 2 could solve this problem in much simpler way. It could also could allow brining final focusing quadrupoles closer to the collision points and reduce $β^*$. It worth noting that many solenoidal detectors introduce rather small rotation angle, *θ<<1*, and compensation could be achieved by a rather weak quadrupoles whose gradients scale as a square root of *θ*.

**Acknowledgements**
Authors are thankful to Nikolay A. Vinokurov, Igor Ya. Protopopov and Yuri M. Shatunov (all Budker Instutute of Nuclear Physics), whose stimulating and fruitful discussion and advices were critical for writing our original paper. We are also thankful to our colleagues around the world who successfully implemented our schemes in operational accelerators. Vladimir Litvinenko is thankful to Joerg Kewisch (Brookhaven National Laboratory) for pointing out and sharing a copy of our preprint translation at DESY.

## Appendix 1. Direct proof of eq. (17)

For the system drown in Fig. 1, the total transport matrix from the entrance to the first solenoid to the exit from the second one is an ordered product of matrices of solenoids $\mathbf{M}_{S1}, \mathbf{M}_{S2}$ and the quadrupole beamline $\mathbf{M}_{QBL}$:

$$\mathbf{M} = \mathbf{M}_{S2}\mathbf{M}_{QBL}\mathbf{M}_{S1}, \tag{A1.1}$$

The quadrupole beamline transport matrix is product of 13 block-diagonal matrices

$$M_{QBL} = M_{D7}M_{Q6}M_{D6}M_{Q5}...M_{Q2}M_{D2}M_{Q1}M_{D1};$$

$$M_{Qi} = \begin{pmatrix} M_{Qix} & 0 \\ 0 & M_{Qiy} \end{pmatrix}; \quad M_{Do} \begin{pmatrix} M_{di} & 0 \\ 0 & M_{di} \end{pmatrix}; \tag{A1.2}$$

and because block diagonal matrices comprise the closed group

$$\alpha \begin{pmatrix} A_1 & 0 \\ 0 & B_1 \end{pmatrix} + \beta \begin{pmatrix} A_2 & 0 \\ 0 & B_2 \end{pmatrix} = \begin{pmatrix} \alpha A_1 + \beta A_2 & 0 \\ 0 & \alpha B_1 + \beta B_2 \end{pmatrix};$$

$$\begin{pmatrix} A_1 & 0 \\ 0 & B_1 \end{pmatrix} \begin{pmatrix} A_2 & 0 \\ 0 & B_2 \end{pmatrix} = \begin{pmatrix} A_1 A_2 & 0 \\ 0 & B_1 B_2 \end{pmatrix}; \tag{A1.3}$$

there product is the block-diagonal matrix:

$$M_{QBL} = \begin{pmatrix} M_x & 0 \\ 0 & M_y \end{pmatrix}. \tag{A1.4}$$

Finally, we can use the fact that solenoid matrix commute with that of rotation and according to (14) can be written in two forms:

$$M_S = R \begin{pmatrix} F & 0 \\ 0 & F \end{pmatrix} \equiv \begin{pmatrix} F & 0 \\ 0 & F \end{pmatrix} R \tag{A1.5}$$

Using left and right forms of (A5) for two solenoids, we have easy proof of eq. (17):

$$M_{S2} = R \begin{pmatrix} F & 0 \\ 0 & F \end{pmatrix}; \quad M_{S1} = \begin{pmatrix} F & 0 \\ 0 & F \end{pmatrix} R;$$

$$M = M_{S2} M_{QBL} M_{S1} = R \begin{pmatrix} FM_x F & 0 \\ 0 & FM_y F \end{pmatrix} R; \tag{A1.6}$$

$$A = FM_x F; \quad B = FM_y F.$$

**Appendix 2. Proof of eq. (20)**

Proof that

$$R \begin{pmatrix} A & 0 \\ 0 & B \end{pmatrix} R = \begin{pmatrix} A_1 & 0 \\ 0 & B_2 \end{pmatrix} \tag{A2.1}$$

requires $A = -B$ comes from simple matrix multiplication:

$$\begin{pmatrix} I\cos\theta & I\sin\theta \\ -I\sin\theta & I\cos\theta \end{pmatrix} \begin{pmatrix} A & 0 \\ 0 & B \end{pmatrix} \begin{pmatrix} I\cos\theta & I\sin\theta \\ -I\sin\theta & I\cos\theta \end{pmatrix} =$$

$$\begin{pmatrix} A\cos^2\theta - B\sin^2\theta & (A+B)\cos\theta\sin\theta \\ -(A+B)\cos\theta\sin\theta & B\cos^2\theta - A\sin^2\theta \end{pmatrix} = \begin{pmatrix} A_1 & 0 \\ 0 & B_2 \end{pmatrix}; \tag{A2.2}$$

which requires

$$(\mathbf{A}+\mathbf{B})\sin 2\theta = 0$$

e.g. for $\sin 2\theta \neq 0$ we need $\mathbf{A}+\mathbf{B}=0$. Naturally resulting matrix has a very simple form:

$$\mathbf{R}\begin{pmatrix} \mathbf{A} & 0 \\ 0 & -\mathbf{A} \end{pmatrix}\mathbf{R} = \begin{pmatrix} \mathbf{A} & 0 \\ 0 & -\mathbf{A} \end{pmatrix}. \tag{A2.3}$$